\documentclass[conference]{IEEEtran}
\IEEEoverridecommandlockouts
\usepackage{cite}
\usepackage{amsmath,amssymb,amsfonts}
\usepackage{algorithmic}
\usepackage{graphicx}
\usepackage{textcomp}
\usepackage{xcolor}
\usepackage{braket}

\usepackage[absolute,showboxes]{textpos}
\usepackage{algorithm}

\newcommand{\copyrightstatement}{
    \begin{textblock}{12.9}(1.6,14.8)    
         \noindent
         \footnotesize
         \centering
         \\
         © 2024 IEEE.  Personal use of this material is permitted.  Permission from IEEE must be obtained for all other uses, in any current or future media, including reprinting/republishing this material for advertising or promotional purposes, creating new collective works, for resale or redistribution to servers or lists, or reuse of any copyrighted component of this work in other works.
    \end{textblock}
}

\def\BibTeX{{\rm B\kern-.05em{\sc i\kern-.025em b}\kern-.08em
    T\kern-.1667em\lower.7ex\hbox{E}\kern-.125emX}}

\begin{document}

\copyrightstatement

\title{Simulator Demonstration of Large Scale Variational Quantum Algorithm on HPC Cluster}

\author{\IEEEauthorblockN{Mikio Morita}
\IEEEauthorblockA{\textit{Quantum Laboratory} \\
\textit{Fujitsu Ltd.}\\
Kawasaki, Japan \\
morita.mikio@fujitsu.com}
\and
\IEEEauthorblockN{Yoshinori Tomita}
\IEEEauthorblockA{\textit{Quantum Laboratory} \\
\textit{Fujitsu Ltd.}\\
Kawasaki, Japan \\
yoshin-t@fujitsu.com}
\and
\IEEEauthorblockN{Junpei Koyama}
\IEEEauthorblockA{\textit{Quantum Laboratory} \\
\textit{Fujitsu Ltd.}\\
Kawasaki, Japan \\
koyama.junpei@fujitsu.com}
\and
\IEEEauthorblockN{Koichi Kimura}
\IEEEauthorblockA{\textit{Quantum Laboratory} \\
\textit{Fujitsu Ltd.}\\
Kawasaki, Japan \\
k.kimura@fujitsu.com}

}

\maketitle

\begin{abstract}
Advances in quantum simulator technology is increasingly required because research on quantum algorithms is becoming more sophisticated and complex.
State vector simulation utilizes CPU and memory resources in computing nodes exponentially with respect to the number of qubits; furthermore, in a variational quantum algorithm, the large number of repeated runs by classical optimization is also a heavy load.
This problem has been addressed by preparing numerous computing nodes or simulation frameworks that work effectively.
This study aimed to accelerate quantum simulation using two newly proposed methods:
to efficiently utilize limited computational resources by adjusting the ratio of the MPI and distributed processing parallelism corresponding to the target problem settings and to slim down the Hamiltonian by considering the effect of accuracy on the calculation result.
Ground-state energy calculations of fermionic model were performed using variational quantum eigensolver (VQE) on  an HPC cluster with up to 1024 FUJITSU Processor A64FX connected to each other by InfiniBand; the processor is also used on supercomputer Fugaku.
We achieved 200 times higher speed over VQE simulations and demonstrated 32 qubits ground-state energy calculations in acceptable time.
This result indicates that > 30 qubit state vector simulations can be realistically utilized to further research on variational quantum algorithms.
\end{abstract}


\section{Introduction}
Quantum computer has received much attention from both hardware and algorithmic perspectives in the last decades.
This is because it was expected to perform faster than classical computation in certain calculation area \cite{qcom}.
Previous studies have reported several experiments of quantum computation conducted on a noisy intermediate scale quantum computer (NISQ\cite{nisq})\cite{exp1,exp2}.
However, the limited capability of NISQ, derived from hardware noise and short coherent time, makes it challenging to validate sophisticated and complex quantum algorithms that contain many qubits or gate operations.
Therefore, quantum simulations on classical computers, which simulate ideal noise less or manually controllable noise environment, are increasingly important for the advancement of quantum algorithm study.

Quantum simulation requires large computational resources of classical computers.
A state vector simulator, for example, Qulacs\cite{qulacs}, must allocate $2^{n+4}$bytes to store $2^{n} $ double-precision complex numbers for representing $n$ qubit quantum state. Consequently, memory resource requirement increases exponentially.
Accordingly, the gate operation must control exponentially increasing computational bases, hence computation costs also increase exponentially.
In addition to the state vector, other quantum simulation methods have been proposed such as tensor network\cite{tensor} and decision diagram\cite{dd}, which also require large computational resources.

The variational quantum algorithm is one of  practical quantum algorithms for NISQ.
For each purpose, they are defined in the form of variational quantum eigensolver (VQE)\cite{vqe1,vqe2,vqe3, vqe4}, quantum approximate optimization algorithm (QAOA)\cite{qaoa1}, and quantum machine learning (QML)\cite{qml1,qml2}.
Since they perform variational calculations by optimizing parameters embedded in the quantum circuit, the quantum circuit must be executed many times in the algorithm.
VQE is an algorithm for quantum chemical calculations that has been studied by simulation and a few qubit real experiments\cite{exp1,exp2}.
The VQE can calculate, in typical usage, ground-state energy or excitation energy\cite{excited1,excited2,excited3} by representing molecular orbitals in the quantum circuit.

Several simulations were performed by previous works.
As a partial example of VQE, the ground-state calculations of BeH$_2$ of 14 qubits\cite{vqesim1}, N$_2$ of 16 qubits\cite{vqesim2},  and naphthalene of 20 qubits\cite{vqesim3} have been reported. 
These problem targets are all fermionic model Hamiltonian.
On the other hand, Heisenberg model Hamiltonian, a limited model interacting with adjacent spins, has been reported up to 40 qubits \cite{40q}.
Similarly, non-variational quantum computational simulations on HPC cluster have been reported.
Quantum simulations up to 72 qubits supported by NASA HPC cluster $Pleiades$ and  $Electra$\cite{hpc1} and 121 qubits on the supercomputer $summit$\cite{hpc2} have been carried out. 
However, research on the large-scale implementation of variational quantum algorithms is still short; thus discussion on the feasibility of this area is important.
Especially, there is a lack of VQE to deal with fermionic models, which require a large number of Hamiltonian terms.

This paper reports the demonstration of fermionic model VQE simulations up to 36 qubits using two newly proposed techniques.
The first technique is to efficiently combine MPI parallel and distributed processing corresponding to the target problem.
By optimizing the node usage ratio of two parallelization methods for the prepared computing nodes, the computing capability can be improved.
The second technique is to slim down Hamiltonian to speed-up the expectation value calculation of quantum simulation; this accelerates VQE while maintaining nearly the same calculation accuracy.
The simulation was carried out on an HPC cluster with up to 1024 FUJITSU Processor A64FX (A64FX) connected to each other by InfiniBand; the processor is also used on supercomputer $Fugaku$.

For the 28- and 32-qubit problem, we demonstrated complete simulation until the obtained energy was converged.
For the 36-qubit problem, only one iteration was performed; the characteristic was revealed.

\section{Methodology}
\label{sec:methods}
In this section, we introduce our methods of VQE simulation on an HPC cluster system. 
The section is divided into the following:

\begin{enumerate}
\item[A.] Computing node
\item[B.] VQE algorithm
\item[C.] MPI and distributed processing
\item[D.] Hamiltonian terms cutoff
\end{enumerate}

Subsections A and B describe known techniques, where A shows the hardware usage of the HPC cluster system and B shows the VQE algorithm.
Subsections C and D describe the newly proposed methods; C explains a way to combine MPI and distributed processing parallelization, and D explains how to accelerate expectation value calculation by slimming down Hamiltonian.
For another method related to coding, we used pySCF\cite{pyscf} for quantum chemical calculations, openfermion\cite{openfer} for hamiltonian qubit mapping, and Qiskit SLSQP for parameter optimization.  Qiskit SLSQP function is based on scipy\cite{scipy}.

\subsection{Computing node}
Large-scale quantum simulations have been performed on a number of multinode HPC cluster systems as in previous studies\cite{hpc1,hpc2}.
Our system consists of FUJITSU PRIMEHPC FX700 (FX700), a computing node with A64FX and 32-GiB memory.
A64FX, the Armv8.2-A instruction set architecture, is a processor also used in the $Fugaku$ supercomputer.
In total, 1024 nodes are connected by InfiniBand EDR.
See Table. 1 for detailed specifications of our HPC cluster system.
Our system is also an HPC cluster as in previous studies; however it was designed for state vector simulator.

Several quantum simulation frameworks such as Qiskit Aer, Intel-QS\cite{intelQS}, and Qulacs\cite{qulacs} have been published.
We used mpiQulacs\cite{mpiQulacs}, an extended Qulacs for distributed simulation, to run on this cluster system.
MpiQulacs was selected because Qulacs is one of the fastest state vector simulation frameworks.

MpiQulacs takes an MPI approach similar to intel-QS\cite{intelQS}, in which the vector representing the quantum state is divided by the number of MPI processes.
The memory used per node is $\frac{2^{n+4}}{p}=2^{n-\log_2 p+4}$ bytes, when the number of MPI processes is $p$ and the number of qubits is $n$. 
Since the FX700 node has 32-GiB memory, considering the amount of memory required for operating systems, python code, and so on, $2^{30+4}=$16 GiB can be used to store quantum states.
Therefore, the maximum number of qubits that can be executed by a single node is up to 30, and the minimum number of $2^ {(n-30)}$ nodes are required when calculating $>$ 30 qubits.
For example, 36-qubit computation requires at least $2^{36-30}=$64 nodes connected by MPI communication.
Refer to the citation\cite{mpiQulacs} for more information on MPI parallel techniques using mpiQulacs.

\begin{table}
\caption{\textbf{HPC Cluster System Specification}}
\label{table1}
\setlength{\tabcolsep}{3pt}
\begin{tabular}{|p{115pt}|p{115pt}|}
\hline
Cluster node&
FUJITSU PRIMEHPC FX700 \\
\# of nodes&
1024 \\
Theoretical peak FLOPS&
3146 TFLOPS (double precision) \\
\hline
\hline
CPU&
FUJITSU Processor A64FX \\
Instruction set architecture&
Armv8.2-A + SVE\\
\# of CPUs per node&
1 \\
\# of cores per node&
48 (computing core)

4 (assistant core)\\
\# of threads per node&
48\\
Base frequency&
2.0 GHz\\
Boost frequency&
2.0 GHz (no boost mode)\\
Theoretical peak FLOPS per CPU&
3.1 TFLOPS (double precision) \\
\hline
\hline
Memory capacity per node&
32 GiB\\
Memory band-width&
1.024 GB/s\\
\hline
\hline
Interconnect&
InfiniBand EDR\\
\hline

\multicolumn{2}{p{210pt}}{An A64FX is installed in a FX700 node.32-GiB memory is stored in the A64FX. More than 1030 FX700 units are actually connected; however, only 1024 units are expected to be used simultaneously. }\\
\end{tabular}
\label{tab1}
\end{table}

\subsection{VQE algorithm}
The VQE algorithm has concerns to be determined carefully, particularly for the design of a quantum circuit ansatz and the overall structure of the algorithm including preprocessing.

To achieve better accuracy in shallower circuits, previous studies have proposed many VQE ansatz. Hardware efficient\cite{exp2}, symmetry preserving\cite{symm}, and UCCSD\cite{uccsd} ansatz are representatives. In addition,  their derivatives include gate fabric symmetry preserving\cite{vqesim3}, UpCCGSD\cite{vqesim2}, jastrow-factor\cite{jastrow} and qCC\cite{qcc}.
Regarding the overall structure of the VQE algorithm, for example, classical calculation for obtaining initial values of variational parameters\cite{accuracy}, adaptive generation of quantum circuits\cite{adapt}, and postprocessing to mitigate hardware noise\cite{ps}, have been proposed as support feature.

The entire algorithm in this study is shown in Fig. 1.
At the beginning of the preprocessing, values of the one- and two-electron integrals were calculated from the chemical problem of interest. 
Then the second quantized Hamiltonian $H$ was generated; it is represented as \eqref{hami0} in the fermionic model.
\begin{equation}H=\sum_{pq}{h_{pq}}{a_p^{\dagger}a_q}+\sum_{pqrs}{h_{pqrs}a_p^{\dagger}a_q^{\dagger}a_ra_s}.\label{hami0}\end{equation}
Values $a^{\dagger}$,$a$, and $h$ mean generation operator, annihilation operator and coefficient, respectively.

These values were sent to two separete flows. One is to generate qubit-mapped Hamiltonians; because our ansatz has the quantum-number-preserving property, the Jordan-Wigner transform method\cite{jw} was chosen instead of Bravyi–Kitaev\cite{bk} or parity basis\cite{parity}. The Hamiltonian $H$ is represented by the sum of Pauli strings, also called observables $P$. 
In \eqref{hami}, \eqref{obse}, and \eqref{sigma}, $w$ represents the observable coefficient, and $\sigma$ is the Pauli matrix of I, X, Y, or Z.
See a citation\cite{ON3} for the details on possible forms of $P$.

\begin{equation}H=\sum{w_i}{P_i}.\label{hami}\end{equation}
\begin{equation}P_i=\bigotimes_{k=1}^n\sigma_k^{(i)}.\label{obse}\end{equation}
\begin{equation}\sigma_k^{(i)}\in \{I,X,Y,Z\}.\label{sigma}\end{equation}

The another is to calculate the configuration interaction singles and doubles (CISD), one of the well-known computational methods of quantum chemistry, to determine the ansatz quantum circuit and initial values of variational parameters. The preprocessing approach to obtain the quantum circuit\cite{qcc,adapt} and the execution of the quantum chemical calculation to generate the initial parameter value\cite{accuracy} have been studied. 
We used the modified version of the methods used in these previous studies by utilizing CISD because of its high accuracy and straightforward correspondence with quantum circuits.
CI coefficients were extracted from the CISD calculation result.
The CI coefficients-generated ansatz $U$ was made up of the product of gate sets, where gate set S represents one-electron excitations on molecular orbitals and D represents two-electron excitations; see \eqref{eq3}. 
Since one CI coefficient corresponds to one excitation operator, it generated one S or D.
Among all the one- or two- electron excitations, the CI coefficient determined the excitation operator to be implemented in ansatz. 
Moreover, the initial parameters of the excitation operator were set, derived from the corresponding CI coefficients.
For S and D, we referred to the gate set of the conventional method\cite{vqesim3}. 
This ansatz form is like a cross between gate fabric symmetry preserving\cite{vqesim3} and UCCSD\cite{uccsd}.
The gate fabric symmetry preserving ansatz is a more advanced form of symmetry preserving, and it has good convergence. In the ansatz in this work, by using the initial value, it converges even better. Therefore, the simulation time should be shorter than the typical ansatz.

After completing the preprocessing, the process moves to the VQE main part.
The ansatz with embedded initial parameters  $U(\theta_0)$ acted through various gate operations on the wave function $\psi_{HF}$ representing the Hartree-Fock state. This created a wave function $\psi(\theta_0)$ that represents the superposition of molecular orbitals.
The parameters were updated by a classical optimization process. This process is repeated until the obtained energy converges to the ground-state energy $E_{gd}$, as shown in \eqref{eq1},\eqref{eq2}.
Types of optimizers include BFGS, Powell, and COBYLA; SLSQP was selected from the viewpoint of the small number of circuit runs until convergence in a noise-less environment.

\begin{equation}U(\theta)=\prod_{pqrs}D_{pqrs}(\theta_{pqrs})\prod_{pq}S_{pq}(\theta_{pq}).\label{eq3}\end{equation}
\begin{equation}\psi(\theta)=U(\theta)\psi_{HF}.\label{eq1}\end{equation}
\begin{equation}E_{gd}=\min_\theta\frac{\bra{\psi(\theta)}H\ket{\psi(\theta)}}{\braket{\psi(\theta)|\psi(\theta)}}.\label{eq2}\end{equation}


\begin{figure*}[t!]
\centerline{\includegraphics{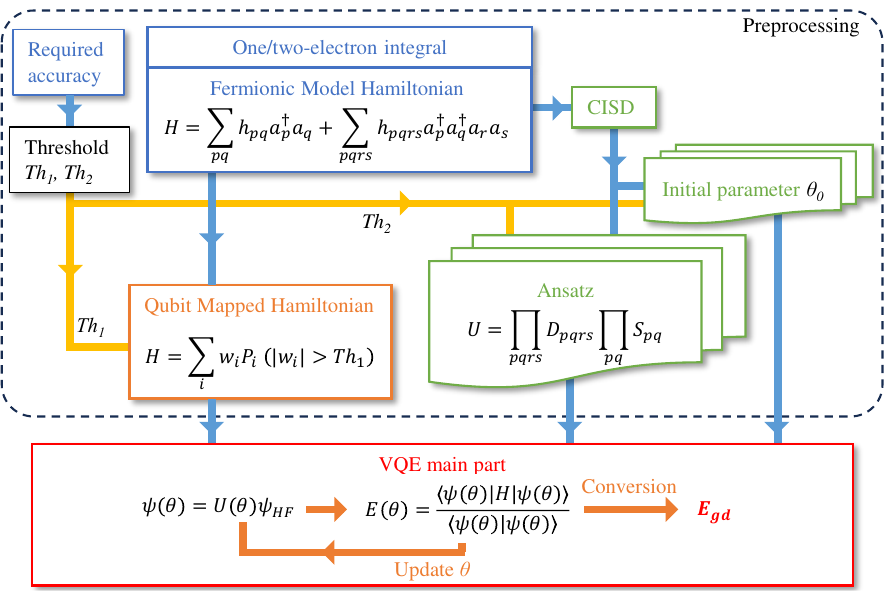}}
\caption{VQE overall algorithm flow. one/two-electron integral and CISD are calculated by pySCF. Threshold $Th_1$ is sent to Hamiltonian output process. $Th_2$ is sent to initial value and ansatz generation. In the VQE main part, the expected values of all observables are summed up to obtain the energy value $E(\theta)$. This is repeated until it converges to the minimum value $E_{gd}$.}\label{fig1}
\label{fig1}
\end{figure*}

The target molecules for ground-state energy calculation were CO$_2$ for 28 qubits and C$_3$H$_6$ for 32 and 36 qubits.
The basis functions are STO-3G for all.

\subsection{MPI and distributed processing}
In general, a method for computing plurality of nodes is not only MPI communication but also distributed processing using remote procedure call (RPC).
RPC is the technology in the connected remote node; the technology can be utilized to speed-up the entire computation by concurrent execution of processes that need not to be executed in series.
Among the widely used RPC frameworks, we used gRPC\cite{gRPC} for its simplicity and high performance.
As an operation form of gRPC, a gRPC client for overall management was assigned to a single node, and numerous gRPC servers for computing were established.

Distributed processing by RPC can speed-up the VQE optimization process because the parameter vector $\theta$ consists of a plurality of scalar parameter $\theta_{pq}$ and $\theta_{pqrs}$ as shown in \eqref{eq3}; thus, for each parameter, the value can be tuned independently on a certain optimizer.
Fig. 2 shows a simple example of two and three gRPC servers with a hundred parameters that achieves 1.96 and  2.83 times increase in speed.
In this case, the parallelization efficiencies are $\frac{102}{2\times 52}\approx0.98$ and $\frac{102}{3 \times 36}\approx0.94$, respectively. 
As a trend, the higher the number of parameters, the better the parallelization efficiency. Variational quantum algorithms are suitable because of its diverse parameters.

When dealing with sequential and parallel processing, the parallelization efficiency is generally derived by Amdahl's law, a famous theory of parallel computational acceleration.
See Amdahl's law speed-up scailing factor \eqref{amd1} and efficiency \eqref{amd2}.

\begin{equation} S_{ideal}=\frac{N_p + N_s}{\frac{N_p}{n_{parallel}} +N_s}.\label{amd1}\end{equation}
\begin{equation} \epsilon_{ideal}=\frac{N_p + N_s}{N_p +N_s n_{parallel}}.\label{amd2}\end{equation}

$S_{ideal}$,$\epsilon_{ideal}$, $N_p$, $N_s$, and $n_{parallel}$ indicate the speed-up scaling factor, parallelization efficiency, number of parallelizable process, sequencial process, and parallelization.
These formulae show that a sequential process becomes a bottleneck where $n_{parallel}$ gets larger.
When distributed processing is performed by HPC cluster, creating a surplus of nodes is possible, as shown in Fig. 2. The efficiency in that case is shown as follows \eqref{eff}.


\begin{figure}[t!]
\centerline{\includegraphics{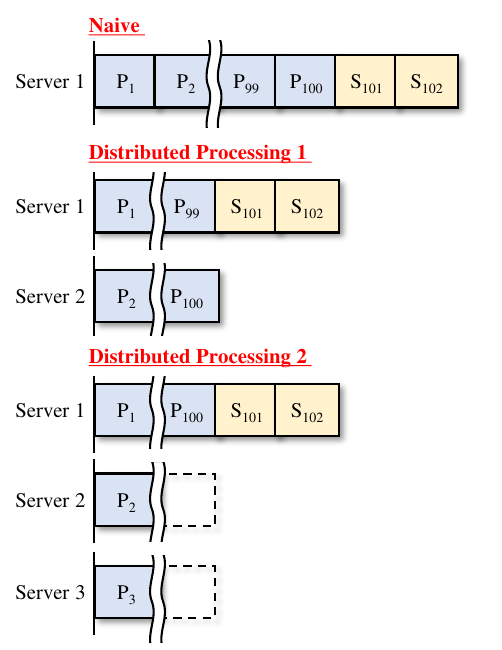}}
\caption{Conceptual diagram of distributed processing on optimization. P$_1$-S$_{102}$ indicates circuit run for one iteration. P$_x$ means a process that can be executed in parallel and S$_x$ means cannot. For example 1 in the middle column, a 1.96 times speedup is achieved and the parallelization efficiency is 98 \%. In the case of example 2 in the lower column, the efficiency is 94 \%.}\label{fig2}
\label{fig2}
\end{figure}

\begin{equation} \epsilon_{DP}=\frac{N_p + N_s}{n_{server} \lceil \frac{N_p}{n_{server}}+ N_s\rceil}.\label{eff}\end{equation}
$\lceil\rceil$ represents the ceiling function, which is the smallest integer greater than value in $\lceil\rceil$.
$\epsilon_{DP}$ and $n_{server}$ indicate the efficiency of distributed processing and the number of gRPC servers, respectively.
$N_p$ and $N_s$ are derived from the number of variational parameters; the relational expression depends on the optimizer. 
For SLSQP we used, $N_p = (\#$ $of$ $parameter)$ and $N_s = 3$.
When the parallelization efficiency of MPI parallel is $\epsilon_{MPI}$, the efficiency of the whole system can be expressed by $\epsilon_{MPI}\epsilon_{DP}$.
If no distributed processing is performed, this efficiency is equal to $\epsilon_{MPI}$.

The proposed method aims to maximize $\epsilon_{MPI}\epsilon_{DP}$ by combining parallel execution by MPI and distributed processing.
As an example of a small 8 nodes system, there are four patterns of node usage. $\times$(number of MPI parallization) - $\times$(number of gRPC server) is written as $\times$8-$\times$1, $\times$4-$\times$2, $\times$2-$\times$4 and $\times$1-$\times$8.
Fig. 3 shows an example.


\begin{figure}[t!]
\centerline{\includegraphics{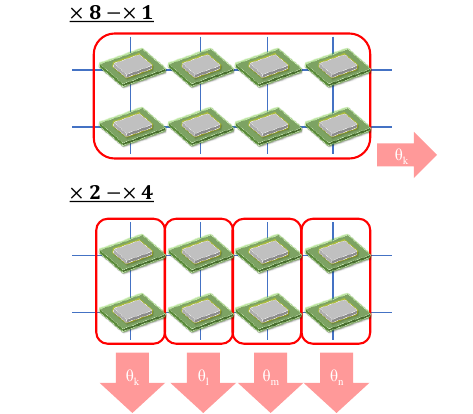}}
\caption{A schematic image of the combination of MPI parallel and distributed processing. The red square frame indicates MPI parallelism. In the image above, eight compute nodes are in MPI connection. In the image below, two nodes with MPI-connected configuration performs four distributed operations. Each configuration is assigned an individual parameter adjustment. The blue line indicates that the node is connected by InfiniBand. They were actually connected through switches, not directly.}\label{fig3}
\label{fig3}
\end{figure}

In our validation, the best combination with a maximum of 1024 nodes was selected. 
Regarding MPI, $\epsilon_{MPI}$ is unpredictable because communication overhead is highly affected.
$\epsilon_{MPI}$ was obtained by executing a quantum circuit once for all MPI patterns. 
In other words, the following configurations were performed:
$\times1-\times1$, $\times2-\times1$, $\times4-\times1$, $\times8-\times1$, $\times16-\times1$, $\times32-\times1$, $\times64-\times1$, $\times128-\times1$, $\times256-\times1$, $\times512-\times1$, $\times1024-\times1$.
Regarding distributed processing, $\epsilon_{DP}$ can be predicted by \eqref{eff}.
Therefore, $\epsilon_{MPI}\epsilon_{DP}$ was figured out.

Distributed processing is also affected by communication overhead, whereas the effect should be negligible for large-scale calculations.
A comparison between the theoretical values and the results obtained in the actual simulation is reported in the section  I\hspace{-1.2pt}I\hspace{-1.2pt}I-B.

\subsection{Hamiltonian terms cutoff}
Calculating the expectation values of the Hamiltonian with the generated wave functions $\bra{\psi}H\ket{\psi}$ must be a major bottleneck in the simulation elapsed time.
The reason is obtaining an expectation value that requires an operations for exponentially increasing computational basis  and the number of observable terms scale $O(n^4)$ in the quantum chemical calculation problem of fermionic model Hamiltonian \cite{ON3}.

The Hamiltonian terms are represented by \eqref{hami}, \eqref{obse}, and \eqref{sigma}.
Since the expectation value of the observables is limited to the range of $-1 \leq \bra{\psi}P\ket{\psi} \leq +1$, it may be useful to cutoff terms with small coefficients to slim down Hamiltonian
because that of the actual term $wP$ moves in the range of $-|w| \leq \bra{\psi}wP\ket{\psi} \leq +|w|$.

In large-scale simulations, knowing the relationship between cutoff threshold, accuracy, and simulation time is diffcult  because the obtained accuracy is not exactly known until the simulation is completed and energy converges to $E_{gd}$.
Our approach determines the Hamiltonian cutoff threshold $Th_1$ by estimating the accuracy from a complete run of  a  relatively small qubit problem.
Algorithm 1 shows psudecode of obtaining Hamiltonian cutoff ratio.

Specifically for 32-qubit problem, the effect on the accuracy was considered by executing the 28-qubit problem simulation untill convergence for several different cutoff ratio.
The cutoff ratio is determined to minimize the number of Hamiltonian terms while satisfying the required accuracy.
Then, moving on to the 32-qubit problem, the cut-off threshold $Th_1$ is determined so that the Hamiltonian term is left by its ratio.

\begin{algorithm}
\caption{Get Hamiltonian cutoff ratio}
\begin{algorithmic} 
\REQUIRE $Smaller \ qubit \ Hamiltonian \ H=\Sigma^N_1 w_i P_i$
\REQUIRE $ a < b \Rightarrow |w_a| \ge |w_b|$
\REQUIRE $Required \ accuracy \ \Delta E$
\STATE $ratio \leftarrow 1$
\WHILE{$ratio \ge 0.1$}
\STATE $E^{ratio}_{gd} \leftarrow executeVQE(H=\Sigma^{(N \times raito)}_1 w_i P_i)$
\IF {$|E_{ratio}-E_{1}| \ge \Delta E$}
\RETURN (ratio \ + \ 0.1)
\ENDIF
\ENDWHILE
\RETURN ratio
\end{algorithmic}
\end{algorithm}

\section{Results}
Data obtained in previous study reveals the property of fermionic model VQE simulation up to 20 qubits \cite{vqesim1,vqesim2,vqesim3}.
In this study, we present the results and characteristics of the VQE simulations up to 36 qubits run on an FX700 HPC cluster system.
As new techniques, we yield parallel execution using a combination of MPI and distributed processing and a Hamiltonian terms cutoff.

This section is divided into the following subsections:

\begin{enumerate}
\item[A.] MPI parallelism
\item[B.] Distrubuted processing
\item[C.] MPI and distributed processing
\item[D.] Hamiltonian terms cutoff
\item[E.] 32-qubit complete VQE simulation
\end{enumerate}

The effects of parallel execution are discussed in terms of MPI parallelism, distributed processing, and their combination.
Then Hamiltonian terms are discussed in subsection D.
Finally, the complete simulation of the 32-qubit problem to convergence is shown.

\subsection{MPI parallelism}
Up to 36-qubit calculation with mpiQulacs has been reported in a previous study \cite{mpiQulacs}.
The study showed the usefulness of quantum computational simulations using MPI for systems with up to 64 A64FX processors.
In this study, we conducted VQE simulations of 28, 32, and 36 qubits with different degrees of MPI parallelism;  furthermore, the elapsed time were obtained.
To save the total running time, the durations of single quantum circuit execution including an expectation value measurement were compared.

Fig. 4(a) shows the relationship between number of the MPI parallelism and execution time.
In the 28-qubit computation, the computation speed decreased when the number of MPI parallelism was increased from 1 to 2. This implies that the communication overhead time exceeded the effect of increasing processing capability.
We barely managed to overcome a single node with over 16 MPI parallelisms.
In other words, single-node computation was very efficient.

For the three datasets, the slopes of the trendlines with more than two parallelisms were almost nearly identical. 
This indicates that the time reduction rate is less associated with the number of qubits in this MPI parallelism region.
The larger the number of MPI parallelisms, the shorter the execution time, with a negative slope.
Since $\frac{\Delta \epsilon_{mpi}}{\Delta n_{parallel}}<0$, doubling the number of MPI parallelisms does not reduce the time in half.
Therefore, the slope was a little gradual.
This can also be seen from the slope of more than 256 parallelisms.

In Fig. 4(b), the speed-up scaling factor from the minimum configuration can be seen. Here, the minimum configuration was $\times1-\times1$ for 28 qubits, $\times4-\times1$ for 32 qubits, and $\times64-\times1$ for 36 qubits.
Theoretically, if there is no communication overhead, the scaling in the $\times512-\times1$ configuration should be larger for those with smaller qubit.
However, in reality, because of communication overhead, single node computations can be faster instead of following the MPI parallel trend.
Accordingly, the scaling of 28 qubits relative to a single node was worse than that of 32 qubits and better than that of 36 qubits.

Fig. 4(c) shows the efficiency $\epsilon_{MPI}$.
The trendline had a negative slope because $\frac{\Delta \epsilon_{mpi}}{\Delta n_{parallel}}<0$.
Since the efficiency was compared from minimum configuration, larger values are naturally obtained in the order of 36, 32, and 28 qubits when comparing on $\times512-\times1$.
For example, 32 qubit was 20 times faster in a $\times512-\times1$ configuration but 128 times more with MPI parallelism, resulting in an efficiency of $\frac{20}{128}\approx0.16$.
A sharp drop in efficiency was commonly observed from more than 256 nodes, suggesting a limit when the number of nodes is increased indefinitely.

The results revealed that MPI parallelism reliably improves the speed by increasing the number of parallelisms, whereas larger ones has a restricted efficiency.


\begin{figure*}[t!]
\centerline{\includegraphics{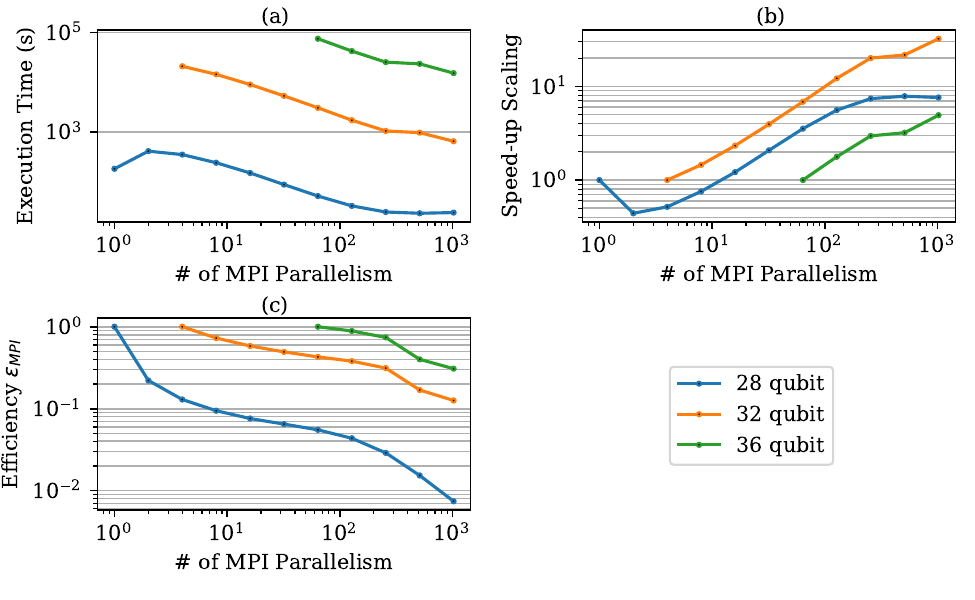}}
\caption{MPI parallel simulation results. The number of parallels is plotted as a power of 2, from 1 to 1024. The three subgraphs are derived from the same result. (a) Elapsed time per quantum circuit execution. (b) Speed-up scailing factor from the minimum configuration. (c) Parallelization efficiency $\epsilon_{MPI}$. }\label{fig4}
\end{figure*}

\subsection{Distrbuted processing}
When gRPC is used for distributed processing, the degree of parallelism is the same as the number of gRPC servers.
A single node gRPC client sent instructions to multiple gRPC servers with up to 1024 nodes. The gRPC server consisted of 1, 4, and 64 nodes for 28-, 32-, and 36-qubit problems, respectively.
As an example, 32 qubit measurements are run in a $\times4-\times N$ configuration where N is the number of gRPC servers.
The distributed processing method ideally has a parallelization efficiency shown in \eqref{eff} however,  a communication overhead should exist between the gRPC client and the gRPC server; the objective is to verify this effect.
Furthermore, this subsection compares the speed-up scaling and the efficiency to MPI parallelism.
Unlike Fig. 4, the comparison was based on the execution time of one iteration; because of taking into account the negative effects of the surplus nodes.

Fig. 5(a) shows the relationship between number of gRPC server and execution time.
The theoretical ideal values are shown as solid lines and the actual values obtained in the simulations are shown as plots.
First of all, the theoretical and the measured values were almost the same. This implies that the communication overhead and the impact of server operations for distributed processing are relatively small.
Second, the execution time trends were similar for the three problem settings. Performance decrease was not observed when the number of nodes was increased from 1 to 2, which was observed in MPI parallel.
We found that the degree of performance improvement by distributed processing was stable.

Fig. 5(b) and (c) display speed-up scaling factor and efficiency $\epsilon_{DP}$ as well as Fig. 4.
The three problem settings showed roughly the same scaling, efficiency.
The subtle difference came from the number of parameters; each had 104, 118 and 153 variational parameters in the circuit.
Efficiency $\epsilon_{DP}$ decreased as the number of gRPC servers increased. This is not primarily due to communication overhead. In the parameter optimization process, parallelizability processes are executed simultaneously on the gRPC servers; as a result, the parallelization efficiency decreases because of the increase in the proportion of the sequential processing part. This is known as Amdal's law.


\begin{figure*}[t!]
\centerline{\includegraphics{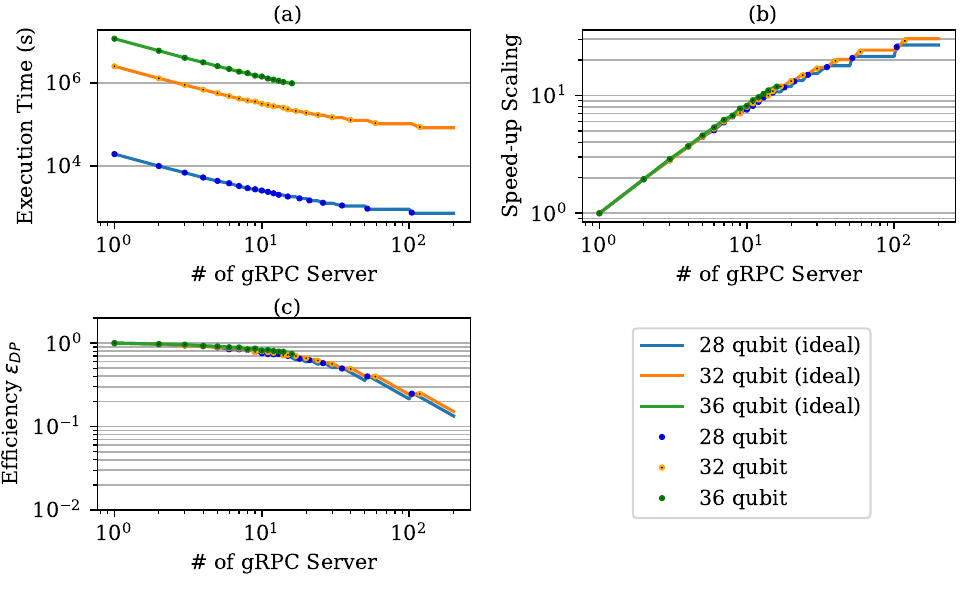}}
\caption{Distrubuted processing simulation results and estimated ideal values. (a),(b) and (c) represents execute time, speed-up scaling and efficiency, respectively as same as Fig. 4. Execute time is during 1 iteration starts to finishes. Plots mean actual obtained values from simulation. Line means ideal expected values drived from \eqref{eff}. }\label{fig5}
\end{figure*}

\subsection{MPI and distributed processing}
Accelerating the simulation by combining MPI and distributed processing in suitable ratio is one of the new attempts of this study; the effect is verified here.
The effects with independent method are discussed in the previous subsections. The question is whether the combination provides improvement.
Since the total number of available nodes is 1024, a combination such as $\times256-\times16$ cannot be configured. Similarly, because a 32 qubits simulation requires at least four nodes of MPI parallel, a configuration like $\times1-\times64$ is not possible.

Fig. 6 shows heatmaps of the improvement in speed and efficiency.
 White color areas are not feasible combinations.
The values on the left and bottom edges of the graph are equal to those obtained in Figs. 4 and 5. Left edge corresponds to independent distributed processing and bottom one to MPI parallel; as the degree of parallelism increases, the degree of speeding-up increases, and the parallelization efficiency deteriorates.
The new results obtained here is provided at the center of the graph. 

Looking at (a), (c) and (e), the maximum speed-up scaling factors of 28, 32 and 36 qubits are found for $\times64-\times16$, $\times128-\times 8$ and $\times128-\times8$, respectively.
While these 3 results agree that using all 1024 nodes is the fastest, and even faster when used in combination rather than independently.
In particular, the 32 qubits results showed a 4-fold speed-up compared to using only the individual methods.

Whole parallelization efficiency $\epsilon_{MPI}\epsilon_{DP}$ is shown in (b), (d), and (f). Efficiency has best value $\epsilon_{MPI}\epsilon_{DP}=1$ for minimum configuration, and tends to get worse as the degree of parallelism increases. 
It is the upper right outermost region, which means the line using 1024 nodes, that is important, where the most efficient combination corresponds to the one with the highest speed improvement represented by (a), (c) and (e).

These results suggest that MPI and distributed processing should be combined for the fastest and most efficient use of prepared nodes.
The results are highly dependent on communication overhead; different hardware communication configurations can lead to different ratio.


\begin{figure*}[t!]
\centerline{\includegraphics{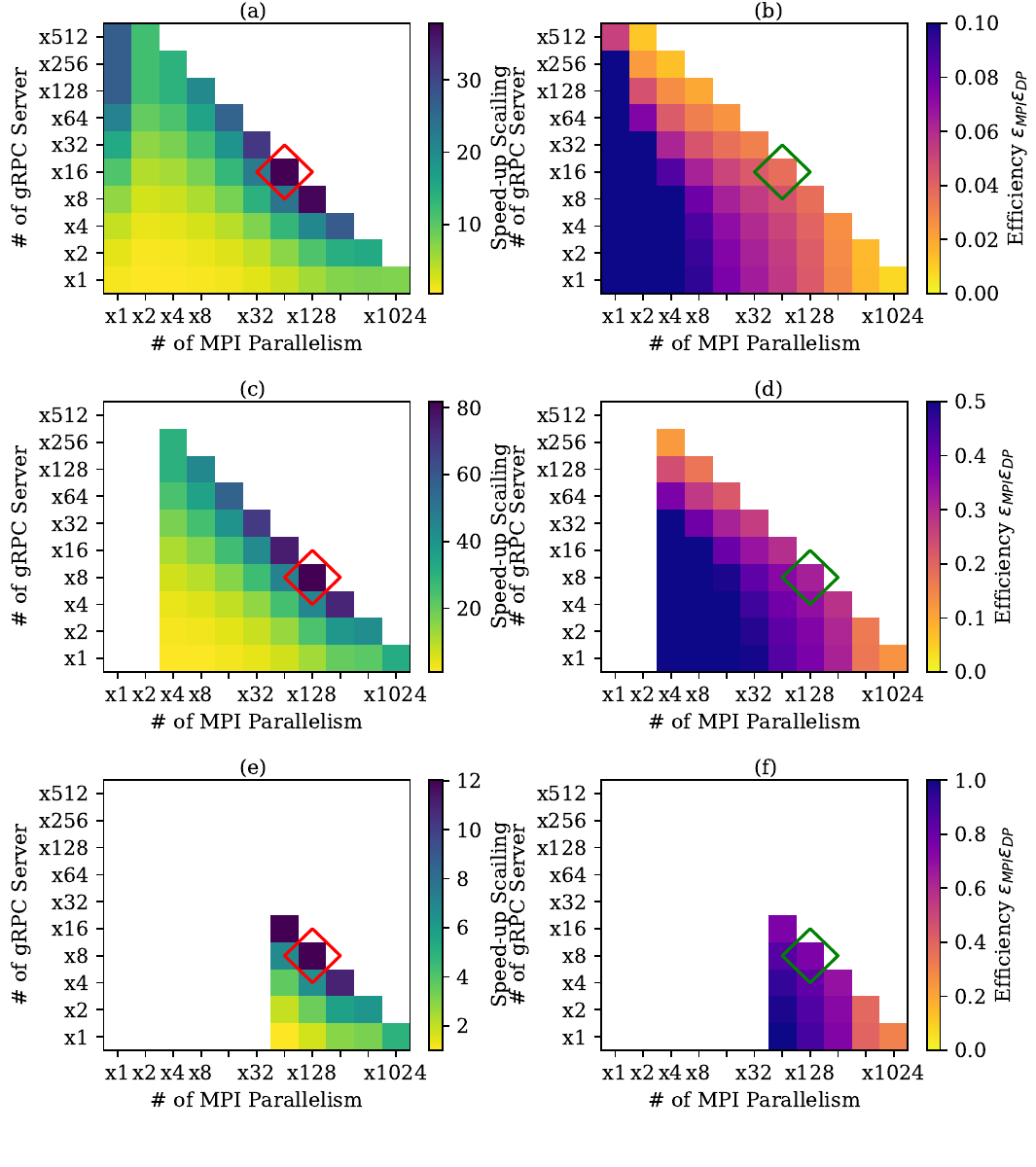}}
\caption{Combination of MPI and distributed processing. (a), (c) and (e) represent speed up scailing of 28, 32 and 36 qubit simulation. (b), (d) and (f) represent parallelization efficiency $\epsilon_{MPI}\epsilon_{DP}$ of 28, 32 and 36 qubit.. The red and green frames in the figure indicate the combination with the highest speed up scaling factor and the most efficient combination when using 1024 nodes. White color areas are not feasible combinations. The speed up scailing and efficiency of the distributed processing are estimated from \eqref{eff}. }\label{fig6}
\end{figure*}

\subsection{Hamiltonian terms cutoff}
Here, we have seen how reducing the number of terms in the Hamiltonian improves the simulation speed and decreases the calculation accuracy.
The Jordan-Wigner transformed Hamiltonian terms were sorted by the absolute values of the coefficients. Then the values below the set threshold $Th_{1}$ were cutoff to slim down and  reconstruct the Hamiltonian.

Fig. 7(a) displays the relationship between the $Th_{1}$ and execution time; the time corresponds to single quantum circuit execution with minimum configuration. The time is normalized to 100\% when the $Th_{1}=0$.
The three colors of plots show similar trends.

Fig. 7(b) displays the relationship between the number of Hamiltonian terms and the execution time. 
A proportional relationship was seen; that can be imagined from the steps of the computation process, calculating the terms individually.

Fig. 7 accounts that adjusting $Th_1$ in the range of $10^{-3}$ can dramatically reduce the time required to obtain the expected value $\bra{\psi}H\ket{\psi}$.

Fig. 8 shows the relationship between the cutoff ratio and the accuracy of the ground-state energy. 
The ratio means that the $Th_1$ was set to cutoff the terms in the particular percentage.
For the 28-qubit problem, accuracy was obtained with cutoff ratios in increments of 10\%. 
For the 32-qubit problem, it was obtained at 0,  60, 70, 80, and 90\% as a reference.
The two problems show the same trend; the accuracy deteriorates as the number of terms to be cutoff increases.
In both problem settings, accuracy is significantly worse when the cutoff ratio is increased to 90\%.

The objective of these measurements were to determine the $Th_1$ value of the complete simulation of 32 qubits. From these results, cutoff ratio was set to 70\%.




\begin{figure}[t!]
\centerline{\includegraphics{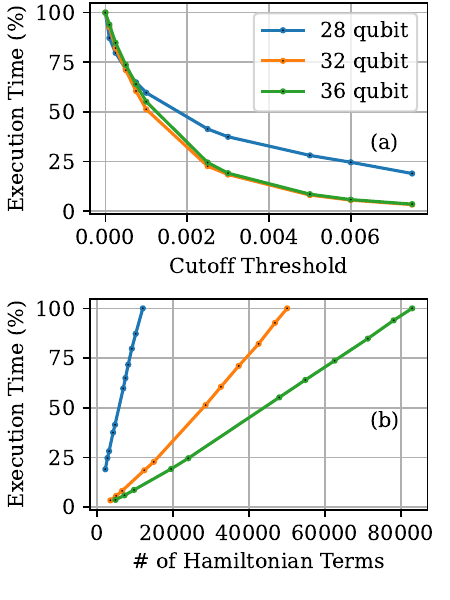}}
\caption{Effect of Hamiltonian term reduction on execution time. The two subgraphs derive from the same result. The time of one quantum circuit execution when the cutoff threshold set to 0, is normalized to 100\%. }\label{fig7}
\end{figure}

\begin{figure}[t!]
\centerline{\includegraphics{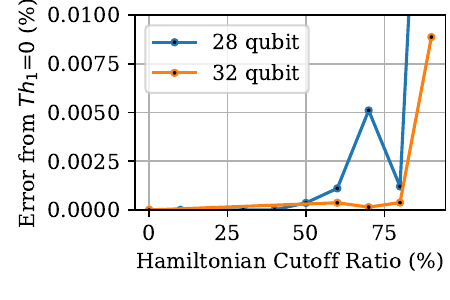}}
\caption{Accuracy and Hamiltonian cutoff ratio. The error indicates the difference from Egd at the threshold $Th_1$=0. The Hamiltonian cutoff ratio indicates the percentage of terms reduced. }\label{fig8}
\end{figure}

\subsection{32-qubit complete VQE simulation}
Large-scale VQE simulations with fermionic Hamiltonian beyond 30 qubits have not been performed; however, we demonstrated VQE simulations at 32 qubits using the introduced techniques.
The demonstration was conducted using the combination of MPI and distributed processing at an optimal ratio and using a Hamiltonian cutoff threshold of 0.0025.
This threshold is set to reduce the Hamiltonian terms by 70\%.
 The accuracy of the ground-state energy is targeted at 0.01 Hartree from that with $Th_1= 0$, hence the error of 0.01\% is acceptable.
Moreover, the execution time for the comparison configuration is also indicated; however, this takes a very long time so that estimation value is shown.
The estimation value was obtained by measuring the time required for one quantum circuit execution and multiplying by the expected number of executions.

Fig. 9 shows the overall execution time of VQE except for preprocessing. 
This is because the time for preprocessing is very short and does not affect the overall time.
The time required by this work, implimenting newly two techniques,  was approximately 15 hours, which is a realistic acceptable time for conducting algorithm research. 
The energy convergence process was shown in Fig. 10 and
the ground-state energies are listed in Table 2. Typical conventional calculation methods are also compared. It means the proper ground-state energy could be obtained by our VQE.

When none or only one of the techniques was applied, the estimated times were 200 days for naive simulation, 3 days for MPI-DP combination, and 40 days for Hamiltonian cutoff, respectively.
The improvement ratio from the naive estimation to this work was about 200. This is somewhat consistent with the approximate percentage improvement $\frac{82}{(1-0.7)} $ derived from discussion in the sectionI\hspace{-1.2pt}I\hspace{-1.2pt}I-C and I\hspace{-1.2pt}I\hspace{-1.2pt}I-D.
The effect of two techniques can be represented approximately simple multiplication because they are independent and do not affect each other.

This result indicates that by implementing the two techniques, the fermionic model VQE simulation until energy convergence can be completed in an acceptable time to proceed with the study of quantum algorithms.
On the other hand, it was estimated that only with a single technique would take more than three days. Increasing the number of nodes from 1024 is another way to solve the problem; however, it requires additional costs.


\begin{figure}[t!]
\centerline{\includegraphics{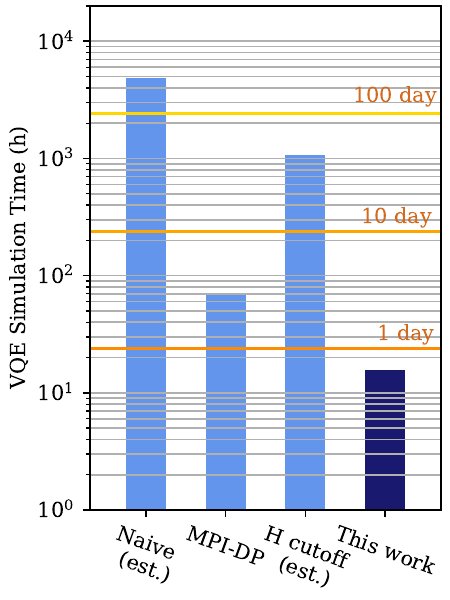}}
\caption{Overall VQE simulation time. The measurement time is only for the VQE main part; preprocessing is excluded. The preprocessing time is relatively small hence negligible. First and third from the left are estimates. The naive estimation done by multiplying the number of executions by the time required for each quantum circuit execution.}\label{fig9}
\end{figure}


\begin{table}
\caption{\textbf{Obtained Ground-state Energy}}
\label{table2}
\setlength{\tabcolsep}{3pt}
\begin{tabular}{p{80pt}|p{95pt}}
\multicolumn{2}{p{210pt}}{28-qubit problem: $CO_2$}\\
\hline
Method&
Obtained ground-state energy (Hartree)\\
\hline
VQE cutoff ratio=90\%&
-185.2966	\\
VQE cutoff ratio=80\%&
-185.2337	\\
VQE cutoff ratio=70\%&
-185.2454	\\
VQE cutoff ratio=60\%&
-185.2380	\\
VQE cutoff ratio=50\%&
-185.2353	\\
VQE cutoff ratio=40\%&
-185.2359	\\
VQE cutoff ratio=30\%&
-185.2359	\\
VQE cutoff ratio=20\%&
-185.2359	\\
VQE cutoff ratio=10\%&
-185.2359	\\
VQE cutoff ratio=0\%&
-185.2360	\\
HF&
-185.0678\\
CCSD&
-185.2698\\
CCSD(T)&
-185.2939\\

\hline
\multicolumn{2}{p{210pt}}{}\\
\multicolumn{2}{p{210pt}}{32-qubit problem: $C_3H_6$}\\
\hline
Method&
Obtained ground-state energy (Hartree)\\
\hline
VQE cutoff ratio=90\%&
-115.7456\\
VQE cutoff ratio=70\%&
-115.7557	 \\
VQE cutoff ratio=0\%&
-115.7559	 \\
HF&
-115.6603\\
CCSD&
-115.8835\\
CCSD(T)&
-115.8848\\

\hline

\multicolumn{2}{p{210pt}}{HF and CCSD mean Hartree-Fock and coupled cluster singles and doubles. Both are typical computational methods for quantum chemical calculations. CCSD(T) is known as a gold standard method.}\\
\end{tabular}
\label{tab2}
\end{table}

\section{Conclusions}
Previous research tried to further develop quantum algorithms by simulating quantum calculations with a large number of qubits.
Despite these were significant steps forward, simulations of variational quantum algorithms, which require multiple quantum circuit executions, had have problem of not completing in an acceptable time.
The reason is, with the number of qubits and variational parameters increases, the number of quantum circuit executions for optimization is high.
Therefore, to the knowledge of authors, VQE for solving the Hamiltonian of fermionic model has only been reported up to 20 qubits\cite{vqesim3}.
In our method, we added newly techniques using the characteristics of the algorithm with parameter optimization and the expectation value calculation.
One is to run the 1024 FX700 nodes not simply in parallel with MPI but to run concurrently through distributed processing. The another is to slim down the Hamiltonian used to calculate the expectation value without losing too much accuracy.

We demonstrated that  32 qubit VQE simulation could be completed in 15 hours.
This is an acceptable time to study the algorithm while changing the gates implemented in ansatz or the problem settings.
Large-scale VQE simulations can be performed by the efficient use of HPC cluster systems.
In addition, two newly techniques were verified.
The computational speed-up due to distributed processing was not significantly affected by the communication overhead.
The behavior followed modified Amdahl's law.
The combination of MPI and distributed processing was found to be faster than using only one method.
In particular, for the 32-qubit problem, the efficient combination gets four times faster than only MPI parallelism or distributed processing.
For the Hamiltonian slimming down, the error was $0.05\%$ even if the terms were reduced by 70\% for the 28-qubit problem.
In verifying VQE by simulation, the result implies that the calculation time can be reduced by several dozen percent by this method if high calculation accuracy is not required.

The 32 qubits simulation demonstration updated the previously reported maximum number of qubits for VQE simulations of fermionic models.
The two newly techniques should also provide improvements not only for VQE but also for other variational algorithms such as QAOA and QML.
Since this work takes the form of statevector simulation, the effect of noise can be also simulated.
Running simulations and experiments on real quantum hardware make it possible to separate the effects of noise from the shortcomings of the algorithm and extract them from the experimental results.

Several things must be considered. For both MPI and distributed processing, communication processing has an effect, so that if the communication environment is not well developed, execution time may increase significantly; and may become unstable.
We chose SLSQP as the optimizer. Other optimization methods may not be as effective as this study if the ratio of sequential processing is large. Conversely, a more parallelism-dominated optimizer could improve further.

This study has allowed us to increase the number of qubits that can be simulated in variational quantum algorithm research.
The techniques will contribute to the practical algorithm search in NISQ.

As a future work, we would like to challenge to complete the VQE simulation with more than 32 qubits in a realistic time. We also would like to expand the scope and work on variational quantum algorithms other than VQE.
Leveraging many nodes at the same time creates a variety of operational problems such as communication. It should be necessary to solve this problem by using general multi-node know-how and unique use of quantum simulation.


\end{document}